\documentclass{article}
\usepackage[a4paper]{geometry}
\usepackage{amssymb}
\usepackage{euscript}
\usepackage{graphicx}
\usepackage{wrapfig}
\usepackage{color}
\usepackage{epsfig}
\usepackage{fullpage}
\def\ts{\theta^s}
\def\ta{\theta_A}
\def\tb{\theta_B}
\def\ct{\cos\vartheta}
\def\st{\sin\vartheta}
\def\cts{\cos^2\vartheta}
\def\sts{\sin^2\vartheta}
\def\cpa{c_{\parallel}}
\def\cpe{c_{\perp}}

\usepackage[colorlinks=true,hypertexnames=true]{hyperref}
\usepackage[style=nature,backend=bibtex,doi=false,isbn=false,url=true,alldates=year]{biblatex}

\begin{document}
\title{NMR in $^3$He-B}
\author{V.~V.~Zavjalov\/\thanks{e-mail: v.zavjalov@lancaster.ac.uk}}

\date{\today}
\maketitle

\begin{abstract}
This text contains a collection of equations useful for understanding
Nuclear Magnetic Resonance (NMR) experiments in superfluid $^3$He-B. This
is a part of my notebook where I try to describe some parts of this
sophisticated system.
\end{abstract}

\section*{Introduction}

Nuclear magnetic resonance (NMR) in superfluid $^3$He is a very powerful
tool used for studying this system since its discovery in 1972.
Superfluidity in fermionic $^3$He is formed via Cooper pairing of atoms, with
pairs having spin 1 and orbital momentum 1. As a result the order
parameter of the system includes both spin and orbital degrees of freedom
and can be written as 3x3 complex matrix. A few different superfluid
phases with different broken symmetries are possible. For the B-phase
degenerate space of the order parameter includes an arbitrary 3D rotation
matrix which describes fixed mutual orientation of spin and orbital
spaces. In NMR experiments we observe motion of this matrix, which means
three degrees of freedom and three spin-wave modes. This is different
from many other magnetic materials where we study motion of magnetization
vector with only two degrees of freedom. Another important feature of
this system is possibility of non-uniform spatial distribution and
topological defects in the order-parameter field (so-called textures).
This can be clearly observed in linear NMR where small oscillations of
the order parameter are happening around the equilibrium texture. One
more feature of this system is spin-orbit interaction. It introduces an
additional non-linear force acting on the order parameter and leads to
important effects such as longitudinal NMR or Homogeneously Precessing
Domain (HPD).

This text contains two almost independent parts. The first one is about
linear NMR. It was written during my work in Vladimir Eltsov's group in
Aalto University (Finland) in 2012-2015. We had a really nice
experimental system for measuring optical magnons trapped in a harmonic
potential formed by the order parameter texture and magnetic field. It was
possible to control shape of the trap and population of individual levels
in it, to observe Bose condensation of magnons, parametric excitation of
other spin-wave modes, interaction of the magnon condensate with texture,
quantized vortices, and free surface of helium. I would like to thank
Petri Heikkinen, Samuli Autti, and Jere M\"{a}kinen who also worked on this
project.

The second part contains Leggett equations for non-linear
NMR and many things related to Homogeneously
Precessing Domain (HPD), a unique coherent
state which is also a very useful tool
for various $^3$He studies. I worked with HPD in Vladimir Dmitriev's
group in Kapitza Institute (Russia) in 2000-2006, and in Pertti Hakonen's
group in Aalto University in 2016-2019.

The text contains many references to Vollhardt and W\"olfle
book~\cite{WV}, usually written as (WV \textless equation
number\textgreater). Many derivations require straightforward but huge
tensor manipulations. A useful tool for this kind of calculations is {\tt
cadabra}~\cite{cadabra}. Some of $^3$He-related parameters and equations
can be found in my {\tt he3lib} library~\cite{he3lib}.

I hope this work will be useful for understanding NMR experiments in $^3$He-B.

\eject
\section*{Part 1. Linear spin waves in $^3$He-B}

At energy scales much less then energy gap~$\Delta$ the state of
superfluid $^3$He-B is described by the order parameter:
\begin{equation}
A_{aj}  = \frac{1}{\sqrt{3}}\ \Delta\ e^{i\varphi} R_{aj}
\end{equation}
where~$\varphi$ is phase, and $R_{aj}$
is a rotation matrix which can be written in terms of rotation axis~$\bf n$
and rotation angle~$\vartheta$ as
\begin{equation}\label{eq1:r_nt}
R_{a j} = \ct\ \delta_{a j} + (1-\ct)\ n_a n_j - \st\ e_{ajk} n_k.
\end{equation}
In small magnetic fields,~$\hbar\gamma H \ll \Delta$ the gap~$\Delta$ and
magnetic susceptibility~$\chi_B$ are isotropic. There are four degrees of
freedom in the system, the phase~$\varphi$, the unit vector~$\bf n$ and
the angle~$\vartheta$. Oscillations of the phase is sound (here we do not
consider normal component and thus have only one sound mode), and
oscillations of the $R_{aj}({\bf n},\vartheta)$ matrix are spin waves. In
the following discussion we are interested only in these three spin-wave
modes. Hamiltonian of the system can be written as a combination of three
components: magnetic energy, energy of spin-orbit interaction and
gradient energy (see VW 6.103, VW 7.17).
\begin{equation}\label{eq1:ham}
\mathcal{H} = F_M + F_{SO} + F_\nabla,
\end{equation}
\begin{eqnarray}
\label{eq1:en_m0}
F_M &=& - ({\bf S} \cdot \gamma {\bf H})
+ \frac{\gamma^2}{2\chi_B}{\bf S}^2,\\
\label{eq1:en_d0}
F_{SO}
&=& g_D\Delta^2 \Big[
           R_{jj}R_{kk}
         + R_{jk}R_{kj}
 - \frac23 R_{jk}R_{jk}\Big]
= g_D\Delta^2 \Big[ R_{jj}R_{kk} + R_{jk}R_{kj}] + \mbox{const.}
,\\
\label{eq1:en_g0}
F_\nabla
&=& \frac12 \Delta^2 \Big[
  K_1 (\nabla_j R_{ak})(\nabla_j R_{ak})
+ K_2 (\nabla_j R_{ak})(\nabla_k R_{aj})
+ K_3 (\nabla_j R_{aj})(\nabla_k R_{ak}) \Big],
\end{eqnarray}
where~$\bf S$ is spin and $\bf H$ is magnetic field.

We are going to write equations of motion for this system. Spin is a not
a canonical variable, its components do not commute with each other and
thus we have to use equations with Poisson brackets. In this approach
evolution of any parameter~$a$ can be written as~$\dot a =
\{\mathcal{H},a\}$. Poisson brackets can be found from microscopic
considerations, from commutation rules in quantum mechanics, or from
symmetry~\cite{1980_pbrackets,2004_bunk_dyn}.

The matrix~$R_{aj}$ for any value of index $j$ can be treated as a
vector in spin space. For such a vector commutation rules can be
written as:
\begin{equation}\label{eq1:brackets_R}
\{S_a, S_b\} = -e_{abc} S_c, \quad
\{R_{aj}, S_b\} = \{S_a, R_{bj}\} = -e_{abc} R_{cj}, \quad
\{R_{aj}, R_{bk}\} = 0.
\end{equation}

It is possible to write equations of motion for the spin~$\bf S$ and
matrix $R_{aj}$ (see later, in the second part of this text), but here we
will use small rotations~$\ts$ of the spin space as coordinates.

\subsection*{Using small rotations $\ts_a$ as coordinates}

We follow derivation of Theodorakis and Fetter
\cite{1983_theodorakis}, see also VW~8.4.6.
Consider some equilibrium distribution of the order parameter matrix
$R^0$. We are going to study small oscillations around the equilibrium.
Arbitrary change of the rotation matrix can be represented as an
additional rotation:
\begin{equation}
R_{a j} = R_{ab}({\bf \ts}) R^0_{bj}.
\end{equation}
Here $\ts$ is a vector and $R_{ab}(\ts)$ is a rotation matrix around the
direction of $\ts$ by angle $|\ts|$. We will use small rotations,
$|\ts|\ll1$. Using formula~(\ref{eq1:r_nt}) for $R_{ab}(\ts)$ one can write
up to second order terms in~$|\ts|$:
\begin{equation}\label{eq1:srot}
R_{a j}
\quad=\quad
\left(
\delta_{ab}
- e_{abc}\ts_c
+ \frac12 \ts_a \ts_b
- \frac12 \delta_{ab}\ts_c \ts_c + o({\ts}^2)
\right) R^0_{bj}
\end{equation}
The Hamiltonian can be written as a function of the spin and angles~$\ts$.
Poisson brackets in the equations of motion can be expanded as
\begin{eqnarray}
\dot S_a
&=& \{\mathcal{H}, S_a\}
\quad=\quad
  \frac{\delta \mathcal{H}}{\delta S_b} \{S_b,S_a\}
+ \frac{\delta \mathcal{H}}{\delta \ts_b} \{\ts_b,S_a\}\\
\dot \ts_a
&=& \{\mathcal{H}, \ts_a\}
\quad=\quad
  \frac{\delta \mathcal{H}}{\delta S_b} \{S_b,\ts_a\}
+ \frac{\delta \mathcal{H}}{\delta \ts_b} \{\ts_b,\ts_a\}
\end{eqnarray}
We are going to obtain linear equations for small~$\ts$. Derivatives of
the Hamiltonian are zero in the equilibrium (energy is in the minimum
at~$\ts=0$) and thus we have only first-order terms there. Thus in the
Poisson brackets we need only zero-order terms. As the matrix $R_{aj}$ is a
function of $\ts$ one can write
\begin{equation}
\{R_{aj}, S_b\} = \frac{d R_{aj}}{d\ts_c}\{\ts_c, S_b\}.
\end{equation}
Using~(\ref{eq1:brackets_R}) and~(\ref{eq1:srot}) one can find commutation
rules for our set of coordinates:
\begin{equation}\label{eq1:brackets_ts}
\{S_a, S_b\} = -e_{abc} S_c, \quad
\{\ts_a, S_b\} = -\{S_a, \ts_b\} = -\delta_{ab}, \quad
\{\ts_a, \ts_b\} = 0.
\end{equation}
It is easy to find the derivative~$\delta \mathcal{H}/\delta S_b$. Using it
and the commutation rules we have:
\begin{eqnarray}
\label{eq1:hameq1_ts}
\dot {\bf S} &=&
  [ {\bf S} \times \gamma {\bf H} ]
- \frac{\delta \mathcal{H}}{\delta {\bf \ts}}\\
\label{eq1:hameq2_ts}
{\bf \dot\ts} &=& \gamma\left(\frac{\gamma}{\chi_B} {\bf S} - {\bf H} \right)
  \quad=\quad \frac{\gamma^2}{\chi_B} {\bf \delta S}
\end{eqnarray}
where we introduced~$\delta {\bf S}$, deviation of the spin from its
equilibrium value ${\bf S^0} = \chi_B{\bf H}/\gamma$. These equations
describe a simple Larmor precession with the additional term
$-\delta\mathcal{H}/\delta\ts$~--- a torque acting on the spin because
of gradient and spin-orbit interactions. This derivative is not trivial,
and will be calculated below.

\subsection*{Variational derivative for small rotations}
Since the energy depends on both angles $\ts_a$ and their gradients
$\nabla_j\ts_a$ we have to use a so-called variational derivative
$\delta\mathcal{H}/\delta \ts_a$~\cite{haar_mech}. An additional
difficulty appears because rotations~$\ts$ do not commute. Let us take
this into account. Using formula~(\ref{eq1:srot}) one can check that two
small successive rotations, $\ta$ and then $\tb$ are equivalent to the
rotation by the angle $\ta + \tb + \frac12 \ta\times\tb$ up to second
order terms in the angles:
\begin{equation}\label{eq1:srot_double}
R_{ac}(\tb)R_{cb}(\ta) \quad=\quad
R_{ab}\left(\ta + \tb + \frac12 [\ta\times\tb] + o(\ta^2,\tb^2)\right).
\end{equation}
Two rotations done in a different order produce a difference $[\ta\times\tb]$.
This can be written as a commutation rule for differentials (see VW~9.14):
\begin{equation}\label{eq1:srot_comm}
\delta\nabla\ts - \nabla\delta\ts = [\delta\ts\times\nabla\ts].
\end{equation}
Consider a small non-uniform rotation, $\delta\ts({\bf r})$,
with boundary condition~$\delta\ts=0$ which changes gradients by $\delta\nabla_j\ts$.
The functional derivative~$\delta\mathcal{H}/\delta\ts_a$ is given by
\begin{equation}
\int_V \frac{\delta\mathcal{H}}{\delta\ts_a} \delta\ts_a\ dr =
\int_V \left(
  \frac{\partial F}{\partial\ts_a}\ \delta\ts_a
+ \frac{\partial F}{\partial\nabla_j\ts_a}\ \delta\nabla_j\ts_a
\right)\ dr
\end{equation}
From mechanical point of view this means that work produced against a
torque~$T_a=-\delta\mathcal{H}/\delta\ts_a$ equals to the total change
of energy. Now we swap differentials~$\delta\nabla\ts$ using the
formula~(\ref{eq1:srot_comm}) and integrate this term by parts using the
zero boundary condition.
\begin{equation}
\int_V \frac{\delta\mathcal{H}}{\delta\ts_a} \delta\ts_a\ dr = 
\int_V \left(
  \frac{\partial F}{\partial\ts_a}\ \delta\ts_a
- \nabla_j\frac{\partial F}{\partial\nabla_j\ts_a}\ \delta\ts_a
+ \frac{\partial F}{\partial\nabla_j\ts_a} e_{abc}\ \delta\ts_b\ \nabla_j\ts_c
\right) dr
\end{equation}
This equation is true for any possible variations~$\delta\ts$. This means that
the integrands are equal at any point and the derivative is:
\begin{equation}\label{eq1:torque}
\frac{\delta\mathcal{H}}{\delta\ts_a} =
 \frac{\partial F}{\partial\ts_a}
- \nabla_j \frac{\partial F}{\partial \nabla_j\ts_a}
+ \frac{\partial F}{\partial\nabla_j\ts_c} e_{abc}\ \nabla_j\ts_b
\end{equation}

\subsection*{Gradient energy}
The gradient energy in $^3$He-B is given by~(\ref{eq1:en_g0}).
Now put the expression for the distorted matrix~(\ref{eq1:srot})
into it. Both original matrix $R^0$ and rotation angles
$\ts$ can be non-uniform here:
\begin{eqnarray}\label{eq1:fgrad}
\frac{2}{\Delta^2}(F_\nabla - F_\nabla^0) 
&=&(2K_1+K_2+K_3)
\ (\nabla_j \ts_a) (\nabla_j \ts_a)
- [K_2 R^0_{aj} R^0_{bk} + K_3 R^0_{ak} R^0_{bj}]
\ (\nabla_j \ts_a) (\nabla_k \ts_b)\\\nonumber
&+& [K_1 R^0_{ak} (\nabla_j R^0_{bk}) + K_2 R^0_{ak} (\nabla_k R^0_{bj}) + K_3 R^0_{aj} (\nabla_k R^0_{bk})]
\ (2e_{abc} + \delta_{ac}\ts_b - \delta_{bc}\ts_a)\ (\nabla_j \ts_c),
\end{eqnarray}
where $F_\nabla^0$~is an energy calculated for the undistorted matrix
$R^0$. Substituting this into~(\ref{eq1:torque}) we have the torque $T_c^\nabla = -\delta F_\nabla/\delta\ts$:
\begin{eqnarray}\label{eq1:tgrad}
\frac{T_c^\nabla}{\Delta^2}
&=& (2K_1+K_2+K_3)\ (\nabla_j \nabla_j \ts_c)
- \nabla_j \left[ (K_2 R^0_{ak} R^0_{cj} + K_3 R^0_{aj} R^0_{ck})\ (\nabla_k \ts_a)\right] \\\nonumber
&+& \frac12 R^0_{ak}\ \left[K_1 (\nabla_j\nabla_j R^0_{bk}) + (K_2 + K_3)( \nabla_j\nabla_k R^0_{bj})\right]
\ (2e_{abc} + \delta_{ac}\ts_b - \delta_{bc}\ts_a)
\end{eqnarray}

If gradients of equilibrium texture $R^0$ can be neglected ("uniform texture") then
\begin{equation}\label{eq1:tgrad_uniform}
T^\nabla_c
=  \Delta^2\ [K\ \delta_{ac}\ \nabla_j\nabla_j - K'\ R^0_{ak} R^0_{cj} \nabla_j\nabla_k]\ \ts_a
\end{equation}
where $K=2K_1+K_2+K_3$ and $K'=K_2+K_3$.

\subsection*{Dipolar energy}

Dipolar energy is given by~(\ref{eq1:en_g0}).
Substituting the small rotation~(\ref{eq1:srot}) we get:
\begin{eqnarray}
R_{jj}R_{kk} + R_{jk}R_{kj}
&=& [ R^0_{jj} R^0_{kk} + R^0_{jk}R^0_{kj} ]\ (1-|\ts|^2)\\ \nonumber
&+& [R^0_{a j} R^0_{kk} + R^0_{a k} R^0_{kj}]\ (\ts_j \ts_a - 2e_{ja b}\ \ts_b) \\ \nonumber
&+& [R^0_{a j} R^0_{a' k} + R^0_{a k} R^0_{a' j}]\ e_{jab}\ \ts_{b}\ e_{ka'b'}\ \ts_{b'}
\end{eqnarray}

Or in terms of axis $\bf n$ and rotation angle $\vartheta$ of matrix $R^0$:
\begin{eqnarray}\label{eq1:edip2}
R_{jj}R_{kk} + R_{jk}R_{kj}
&=& \frac12(4\ct+1)^2  -\frac12\\
&-& 4\st (4 \ct + 1)\ ({\bf n} \cdot {\bf \theta}^s)\nonumber\\ \nonumber
&-& (4\ct+1)(\ct+1)\ |\ts|^2 \\ \nonumber
&+& (9 + 3\ct - 12\cts)\ ({\bf n} \cdot {\bf \theta}^s)^2 \nonumber
\end{eqnarray}

The torque $T_c^D = -\delta F_D/\delta\ts$ is
\begin{eqnarray}\label{eq1:tdip}
\frac{{\bf T}^D}{\Delta^2} = 
&=& 2 g_D (4 \ct + 1) \left[
  2\st\ {\bf n} + (\ct+1)\ {\bf\ts}
  \right] \\ \nonumber
&-& 2 g_D (9 + 3\ct - 12\cts)\ ({\bf n} \cdot {\bf \theta}^s) {\bf n}
\end{eqnarray}


Equilibrium texture stays in the minimum of the dipolar energy
where $\ct=-1/4$. In this case the dipolar torque is
\begin{equation}\label{eq1:tdip2}
{\bf T}^D = - 15 g_D\Delta^2\ ({\bf n\cdot\ts}) {\bf n}
\end{equation}



\subsection*{Spin waves in a uniform texture}

Put the dipolar torque~(\ref{eq1:tdip2}) and the gradient
torque~(\ref{eq1:tgrad_uniform}) into the equation~(\ref{eq1:hameq1_ts}),
differentiate it over time and exclude $\ts$ using the
equation~(\ref{eq1:hameq2_ts}). Here we work only with first order terms
in~$\ts$ and thus do not care about commutation of
derivatives~(\ref{eq1:srot_comm}).
\begin{eqnarray}\label{eq1:ham_eq3}
\delta{\bf \ddot S} &=& [\delta{\bf \dot S}\times \gamma {\bf H}] 
  + {\bf\hat\Lambda}\ \delta{\bf S}, \\\nonumber
&& \mbox{where}\quad  \Lambda_{ab} = \frac{\Delta^2\gamma^2}{\chi_B} \left[
K\ \delta_{ab}\delta_{kj}
- K'\ R^0_{ak}R^0_{bj}\right] \nabla_j\nabla_k
- \Omega_B^2\ n_a n_b.
\end{eqnarray}
Here we introduced Leggett frequency:
\begin{equation}\label{eq1:omega_b}
\Omega_B^2 = \frac{\gamma^2\Delta^2}{\chi_B} 15g_D.
\end{equation}

For a flat wave $\delta {\bf S} = {\bf s} \exp(i\omega t + i{\bf k}{\bf x})$
with frequency~$\omega$ and wave vector~${\bf k}$ in the magnetic field directed along $z$~axis

\begin{equation}
\left[
\left(
\begin{array}{ccc}
\omega^2 & -i\omega\omega_L & 0 \\
i\omega\omega_L & \omega^2 & 0 \\
0 & 0 & \omega^2
\end{array}
\right)
+ {\bf\hat\Lambda}
\right]
\left(
\begin{array}{c}
s_x \\ s_y \\ s_z
\end{array}
\right)
=0,
\end{equation}
\begin{equation}
\qquad
\Lambda_{ab} = \frac{\Delta^2\gamma^2}{\chi_B} \left[
K\ \delta_{ab}{\bf k}^2
- K'\ R^0_{ak}k_k R^0_{bj}k_j\right]
- \Omega_B^2\ n_a n_b.
\end{equation}
where~$\omega_L=\gamma H$. Solution corresponds to zero determinant of the matrix.
Using the fact that ${\bf\hat\Lambda}$ is symmetric this can be written as:
\begin{eqnarray}
[ (\Lambda_{xx} + \omega^2)(\Lambda_{yy} + \omega^2)
  - \Lambda_{xy}^2 - \omega^2\omega_L^2](\Lambda_{zz} + \omega^2)
&+&\\\nonumber
2\Lambda_{xy}\Lambda_{yz}\Lambda_{xz}
 - (\Lambda_{xx} + \omega^2)\Lambda_{yz}^2
 - (\Lambda_{yy} + \omega^2)\Lambda_{xz}^2
&=& 0,
\end{eqnarray}
or
\begin{eqnarray}
\omega^6
&+& \omega^4[ \Lambda_{xx} + \Lambda_{yy} + \Lambda_{zz} - \omega_L^2 ]\\\nonumber
&+& \omega^2[
  \Lambda_{xx}\Lambda_{yy}
+ \Lambda_{yy}\Lambda_{zz}
+ \Lambda_{zz}\Lambda_{xx}
- \Lambda_{xy}^2
- \Lambda_{yz}^2
- \Lambda_{xz}^2
- \omega_L^2\Lambda_{zz}
]\\\nonumber
&+& \Lambda_{xx}\Lambda_{yy}\Lambda_{zz}
+ 2\Lambda_{xy}\Lambda_{yz}\Lambda_{xz}
- \Lambda_{xx}\Lambda_{yz}^2
- \Lambda_{yy}\Lambda_{zx}^2
- \Lambda_{zz}\Lambda_{xy}^2
= 0,
\end{eqnarray}

We have a third-order equation for $\omega^2$. It gives three doubly
degenerate spin-wave modes with positive and negative frequencies. The
equation can be analytically solved for arbitrary wave-vector~${\bf k}$,~${\bf n}$
and~$\omega_L$ in the uniform texture ($\nabla R^0 \ll k$). Calculation
is implemented in {\tt he3lib}~\cite{he3lib}.

\subsection*{Uniform precession}

If we neglect all gradient terms then $\Lambda_{ab} = -\Omega_B^2 n_a n_b$ and
equation~(\ref{eq1:ham_eq3}) is (compare with~\cite{2006_jetpl_catrel}):
\begin{equation}
\omega^6 - \omega^4\big(\omega_L^2 + \Omega_B^2\big) + \omega^2\omega_L^2\Omega_B^2 n_z^2 = 0.
\end{equation}
NMR experiments are often done at a fixed frequency $\omega$. We can solve the equation
for $\gamma H = \omega_L$ and find magnetic field where the resonance is observed:
\begin{equation}
\gamma H  =  \omega \sqrt{ \frac{\omega^2 - \Omega_B^2}{\omega^2 - \Omega_B^2 n_z^2} }.
\end{equation}
At a fixed magnetic field we solve the equation for~$\omega$ and find three spin-wave modes:
\begin{equation}
\omega^2 = \frac12(\omega_L^2 + \Omega_B^2) \pm \sqrt{\frac14(\omega_L^2 + \Omega_B^2)^2 - \omega_L^2\Omega_B^2 n_z^2},
\qquad \omega^2 = 0
\end{equation}
For $\gamma H \gg \Omega_B$ we have a well-known expressions for transverse and longitudinal NMR frequency:
\begin{equation}
\omega = \gamma H + \frac{\Omega_B^2}{2\gamma H}\ (1-n_z^2),
\qquad
\omega = \Omega_B\ n_z,
\qquad
\omega = 0,
\end{equation}
Here meaning of $\Omega_B$ becomes clear, it is frequency of the
longitudinal NMR in a texture with~$\bf n || H$.

\subsection*{Separating equations for transverse and longitudinal magnons}

Let's use complex {\it spherical coordinates}.
For an arbitrary vector $A$:
\begin{equation}
A_{\pm} = \frac{1}{\sqrt2} (A_x\pm i A_y),\quad
A_0 = A_z,\qquad
A_i B_i = A_q B^*_{q},\quad\mbox{where $q=0,+,-$}
\end{equation}
In terms of polar and azimuthal angles $\alpha$ and $\beta$:
\begin{equation}
A_\pm = \frac{|A|}{\sqrt2}\sin\beta\exp(\pm i\alpha),\qquad
A_0 = |A|\cos\beta
\end{equation}

If field $\bf H$ is directed alone the $z$ axis, then $[A \times H]_q = -q i A_q H$.
Now one can write the equation in spherical coordinates for a harmonic
oscillation $\delta S_q = s_q e^{i\omega t}$:
\begin{equation}
\left[
\frac{\gamma^2\Delta^2}{\chi_B}
  \big[-K\ \delta_{qp} \nabla_j\nabla_j
 + K'\   R^0_{qj}R^0_{p^*k} \nabla_j\nabla_k\big]
+ \Omega_B^2\ n_q n_p^*
\right] s_p =
\omega(\omega-q\gamma H) s_q
\end{equation}

Now let's remove non-diagonal coupling between different modes. According
to~\cite{1983_theodorakis} this coupling is small in the high-frequency
limit $\omega\gg\Omega_B$ (as $\Omega_B^2/\omega^2$). Then
we can write (without summation of the $q$ index):
\begin{equation}\label{eq1:spinwaves1}
\left[
\frac{\gamma^2\Delta^2}{\chi_B}
  \big[-K\ \nabla_j\nabla_j
 + K'\  R^0_{qj}R^0_{q^*k} \nabla_j\nabla_k\big]
+ \Omega_B^2\ |n_q|^2
\right] s_q =
\omega(\omega-q\gamma H) s_q
\end{equation}

Now we can write separate equations for transverse ($q={+},{-}$) and
longitudinal ($q=0$) modes. We use the fact that
\begin{equation}
R^0_{+j} R^0_{-k}\label{eq1:rr}
= \frac12 (R^0_{xj}R_{xk}+R^0_{yj}R_{yk})
= \frac12(\delta_{jk} - R^0_{zj}R_{zk}),
\end{equation}
and $R^0_{zj}$ is an equilibrium orbital anisotropy
axis: $\hat L^0_j = R^0_{aj} \hat S^0_a = R^0_{zj}$.

\begin{eqnarray}\label{eq1:tr_spinwaves}
\Big[
 - c_\perp^2\ \nabla^2
 - (c_\parallel^2-c_\perp^2) \hat L^0_j \hat L^0_k \nabla_j\nabla_k
+ \frac12 \Omega_B^2 \sin^2\beta_N
\Big] s_+ &=&
\omega(\omega-\gamma H)\ s_+\\
\Big[\label{eq1:lo_spinwaves}
 - C_\perp^2 \nabla^2
 - (C_\parallel^2-C_\perp^2) \hat L^0_j \hat L^0_k \nabla_j\nabla_k
+ \Omega_B^2 \cos^2\beta_N
\Big] s_0 &=&
\omega^2\ s_0
\end{eqnarray}
here we introduce parameters:
\begin{equation}
c_\parallel^2 = C_\perp^2 = \frac{\gamma^2\Delta^2}{\chi_B} K,\quad
c_\perp^2 = \frac{\gamma^2\Delta^2}{\chi_B}(K-K'/2),\quad
C_\parallel^2 = \frac{\gamma^2\Delta^2}{\chi_B}(K-K'),\qquad
\end{equation}

We have skipped the equation for~$s_{-}$. It can be obtained from one for~$s_{+}$ by changing
$\omega$ by $-\omega$ and thus have same solutions but with an opposite sign. As expected we
have three doubly degenerate modes with positive and negative~$\omega$.
Note that these equations have been obtained in the assumption $\omega\gg\Omega_B$.

\subsection*{Spin-wave velocities in a weak coupling approximation}

Gradient energy coefficients
have been obtained in~\cite{1975_cross} (see also VW~7.25).
\begin{equation}
2 K_1 + K_2 + K_3 = -\frac{2}{\Delta^2}\left(\frac{\hbar}{2m}\right)^2 (4+\delta)c,
\qquad
K_2 = -\frac{2}{\Delta^2}\left(\frac{\hbar}{2m}\right)^2 c,
\qquad
K_3 = -\frac{2}{\Delta^2}\left(\frac{\hbar}{2m}\right)^2 (1+\delta)c
\end{equation}
where
\begin{equation}
c=-\frac{\rho_s}{10}
\ \frac{3+F_1^a}{3+F_1^s}
\ \frac{1}{1+F_1^a(5-3\rho_s/\rho)/15}
,\qquad
\delta = \frac{F_1^a \rho_s/\rho}{3+F_1^s(1-\rho_s/\rho)}
\end{equation}
Then
\begin{equation}\label{eq1:swvel_theor}
c_\parallel^2 = C_\perp^2 =
-\frac{2\gamma^2}{\chi_B}
\left(\frac{\hbar}{2m}\right)^2
(4+\delta)c
,\quad
c_\perp^2 =
-\frac{2\gamma^2}{\chi_B}\left(\frac{\hbar}{2m}\right)^2
(3+\delta/2)c
,\quad
C_\parallel^2 =
-\frac{2\gamma^2}{\chi_B}\left(\frac{\hbar}{2m}\right)^2
2c
\end{equation}

Without Fermi liquid corrections $K_1=K_2=K_3$ and
\begin{equation}
c_\perp/c_\parallel = \sqrt{3/4},\qquad
C_\perp/C_\parallel = \sqrt{2}.
\end{equation}
In many papers one of these two sets of spin-wave velocities is usually
used. For example, in early Leggett's papers it is $C_\perp/C_\parallel$~\cite{1975_he3_teor_leggett},
in Fomin's papers it is $c_\perp/c_\parallel$~\cite{1980_jetp_fomin_spinwaves}.

\subsection*{Motion of $\bf L$, $\bf n$ and $\vartheta$ in the spin wave}

\def\dt{\delta\vartheta}
\def\dn{\delta n}

A common question is how $\bf L$, $\bf n$ and $\vartheta$ move in the
transverse NMR. Here we calculate this explicitly. First let us find
deviations of rotation angle and axis of the matrix $R_{aj}$, ~$\bf\dn$
and~$\dt$ caused by rotation $\bf\ts$. In a linear approximation:
\begin{equation}
R_{aj} =
R^0_{aj}
+ \frac{\partial R^0_{aj}}{\partial\vartheta}\dt
+ \frac{\partial R^0_{aj}}{\partial n_k}\dn_k
= (\delta_{ab} - e_{abc}\ts_c) R^0_{bj}
\end{equation}
Using expression~(\ref{eq1:r_nt}) for~$R^0_{aj}$ and finding convolutions
with~$\delta_{aj}$, $n_a$ and $n_j$ one can find:



\begin{equation}\label{eq1:dtdn}
\dt = {(\bf n \cdot \ts)}
,\qquad
{\bf \dn} =
- \frac12\ {[\bf n \times \ts]}
+ \frac{{\bf \ts} - {(\bf \ts\cdot n)}\ {\bf n}}{2 \tan\vartheta/2}
\end{equation}

Orbital and spin anisotropy axes are connected by the order parameter:
$L_j = S_a R_{aj}$. Using~(\ref{eq1:srot}) and~(\ref{eq1:hameq2_ts}) one can write
up to the first order of $\ts$:
\begin{equation}\label{eq1:dl}
L_j =
\frac{\chi_B}{\gamma^2}
(\gamma {\bf H} + \dot \ts - [\gamma {\bf H} \times \ts] )_a R^0_{aj}
\end{equation}

Consider a harmonic transverse spin wave with frequency~$\omega$ in the field
$\bf H\parallel \hat z$. Then
\begin{equation}
\dot\ts = \frac{\gamma^2}{\chi_B} {\bf \delta S}
,\qquad
\ts = - \frac{\gamma^2}{\omega \chi_B} [{\hat z} \times {\bf\delta S}]
\end{equation}
and substituting this to~(\ref{eq1:dtdn}) and~(\ref{eq1:dl}) we write~$\dt$ and~$\bf L$:
\begin{equation}\label{eq1:dtdn_harm}
\dt = - \frac{\gamma^2}{\omega \chi_B} [{\bf n} \times {\bf\delta S}]_z
=  \frac{\gamma^2}{\omega \chi_B} (n_x \delta S_y - n_y \delta S_x)
\end{equation}
\begin{equation}\label{eq1:dl_harm}
L_j =
\frac{\chi_B H}{\gamma} R^0_{zj}
 + \frac{\omega - \gamma H}{\omega}\ \delta S_a R^0_{aj}
\end{equation}

One can see, that the motion of the orbital anisotropy axis $\bf L$ is small if
the precession frequency is close to $\gamma H$, and motion of~$\vartheta$
is is small if the texture is close to the Leggett configuration
($\bf n\parallel \hat z$).

In the equilibrium dipolar energy stays at the minimum, where $\ct=-1/4$
(so-called Leggett angle). In the spin wave it goes out of the minimum
and thus some additional energy appear. It depends on the orientation of~$\bf n$
and thus can modify equilibrium texture. This effect causes self-localization
of magnons~\cite{2012_PPD_Helsinki}. From the expression~(\ref{eq1:edip2}) for the dipolar energy:
\begin{equation}
F_D - F_D^0 = \frac{15}{2}g_D\Delta^2\ {(\bf n \cdot \ts)}^2
 = \frac{15}{2}g_D\Delta^2\ \delta\vartheta^2
\end{equation}
Average change in the energy, caused by the transverse spin wave can
be calculated using~(\ref{eq1:dtdn_harm}):
\begin{equation}
\langle F_D - F_D^0 \rangle =
\frac{15}{4} g_D\Delta^2
\left( \frac{\gamma H}{\omega}\right)^2
\sin^2\beta_N\ \beta_M^2
\end{equation}

Another useful formula is a relation between equilibrium textural angles
$\beta_N$ and $\beta_L$. In the equilibrium $\bf \hat S^0=\hat z$ and
\begin{equation}
\cos\beta_L = \hat L^0_z = R^0_{zz} = \ct + (1-\ct) \cos^2\beta_N.
\end{equation}
In the minimum of the dipolar energy
\begin{equation}\label{eq1:beta_l_n}
\sin^2\frac{\beta_L}{2} = \frac58 \sin^2\beta_N.
\end{equation}

\subsection*{Quasiclassical equation for magnons}

Consider case of short spin waves, in which spin changes on much
smaller distances then the texture. We can represent such a wave as a set
of flat waves $s_q \propto \cos({\bf k}\cdot {\bf r})$ where wave
vector~$k$ also changes slowly. Substituting the flat wave
into~(\ref{eq1:tr_spinwaves}) and~(\ref{eq1:lo_spinwaves}) we get:
\begin{eqnarray}
\label{eq1:tr_spinwaves_qc}
 c_\perp^2\ {\bf k}^2 + (c_\parallel^2-c_\perp^2) ({\bf k}\cdot\hat {\bf L}^0)^2
+ \frac12 \Omega_B^2 \sin^2\beta_N
&=& \omega(\omega-\gamma H)\\
\label{eq1:lo_spinwaves_qc}
 C_\perp^2 {\bf k}^2 + (C_\parallel^2-C_\perp^2) ({\bf k}\cdot\hat {\bf L}^0)^2
+ \Omega_B^2 \cos^2\beta_N &=& \omega^2
\end{eqnarray}

Here meaning of~$c_{\perp,\parallel}$ and~$C_{\perp,\parallel}$ becomes
clear. In a short-wave limit where magnetic field and spin-orbit
interaction are not important ($\omega\gg\Omega_B,\gamma H$) we have a
linear dispersion laws where~$c_{\perp,\parallel}$ are velocities of
transverse waves, propagating perpendicular and parallel to $\hat
{\bf L}^0$ direction; $C_{\perp,\parallel}$ are same velocities for
longitudinal waves.

The first equation describes transverse spin waves, which are similar to
that in other magnetic systems. In the presence of magnetic field it has two
solutions $\omega(k)$, which are called acoustic
(low $\omega$) and optical (high $\omega$) magnons. The second
equation for longitudinal waves is unique for $^3$He.

\subsection*{Schr\"odinger equation for magnons}

Consider a long-wave optical magnons with frequencies
$\omega\approx\gamma H$ in the texture where~$\bf n$ is close to
vertical. In this case we can write equation for transverse spin
waves~(\ref{eq1:tr_spinwaves}) in a form of a Schr\"odinger equation for
magnon quasiparticles, where complex transverse spin $s_{+}$ plays role
of a wave function and precessing frequency $\omega$ plays role of an
energy. Effect of texture on the gradient terms is neglected because
it adds only a small correction to the total gradient energy. The dipolar
energy, which also depends on texture, can be of the same order as the total
gradient energy.
\begin{equation}\label{eq1:schred}
\left[
- \frac{c_\perp^2}{\gamma H}\ (\nabla_x^2+\nabla_y^2)
- \frac{c_\parallel^2}{\gamma H}\ \nabla_z^2
+ \frac{\Omega_B^2}{2\gamma H} \sin^2\beta_N + \gamma H
\right] s_{+} =
\omega\ s_{+}
\end{equation}
\begin{equation}
s_{+}
 = \frac1{\sqrt2}\ \frac{\chi_B H}{\gamma}\sin\beta_M\ e^{i(\alpha_M + \omega t)}
\end{equation}
Both texture (spatial distribution of $\beta_N$) and magnetic field form
a potential for magnons. It our experiment we used a combined effect of
a flare-out texture in the cylindrical cell and non-uniform field of
a small longitudinal coil to create a harmonic energy trap for magnons.


\subsection*{Magnons in a harmonic trap}

Schr\"odinger equation in a cylindrical harmonic potential can be solved analytically.
Let us write it in the form:
\begin{equation}\label{eq1:schred3d}
\left[
- \frac{\hbar^2}{2m_\perp}\ (\nabla_x^2 + \nabla_y^2)
- \frac{\hbar^2}{2m_\parallel}\ \nabla_z^2
+ \hbar\omega_0
+  \frac{m_\perp\omega_r^2}{2}\ r^2 +  \frac{m_\parallel\omega_z^2}{2}\ z^2
\right] s_{+} =
\hbar \omega\ s_{+}
\end{equation}
where parameters~$\omega_0$, $\omega_r$ and~~$\omega_z$ describes the potential and
masses~$m_{\perp,\parallel}$ in the case of magnons are:
\begin{equation}\label{eq1:magnon_mass}
m_{\perp,\parallel} = \frac{\hbar}{2 c_{\perp,\parallel}^2}\gamma H.
\end{equation}

Then eigenvalues are
\begin{equation}
\omega = \omega_0 + \omega_r(2n_r + |n_\phi| + 1) + \omega_z(n_z + 1/2),
\end{equation}
for $n_r,n_z=0,1,2\ldots$, $n_\phi=0,\pm1,\pm2\ldots$,
and  normalized solutions are:
\begin{equation}
s_{+} = s_+^r(r)\ s_+^\phi(\phi)\ s_+^z(z),
\end{equation}
where
\begin{eqnarray}
s_+^r(r) &=&
\frac{1}{a_r} \sqrt{ \frac{2\ n_r!}{(n_r + |n_\phi|)!} }
\ \left(\frac{r}{a_r}\right)^{|n_\phi|}
\exp\left(-\frac{r^2}{2a_r^2}\right)
L_{n_r}^{|n_\phi|}\left(\frac{r^2}{a_r^2}\right)\\
s_+^\phi(\phi) &=&
\ \frac{1}{\sqrt{2\pi}} \exp(i n_\phi \phi)\\
\nonumber
s_+^z(z) &=& 
\ \sqrt{\frac{1}{a_z \sqrt{\pi}\ 2^{n_z}\ n_z!}}
\ \exp\left(-\frac{z^2}{2a_z^2}\right)
H_{n_z}\left(\frac{z}{a_z}\right),
\nonumber
\end{eqnarray}
$L_n^m(x)$ and $H_n(x)$ are Laguerre and Hermite polynomials:
\begin{equation}
L_n^m(x) = \frac{1}{n!}\ x^{-m}\ e^x\ \frac{d^n}{dx^n} (x^{n+m}\ e^{-x}),\qquad
H_n(x) = (-1)^n\ e^{x^2}\ \frac{d^n}{dx^n}\ e^{-x^2},
\end{equation}
and $a_r$ and $a_z$ are sizes of the wave:
\begin{equation}\label{eq1:wave_size}
a_r=\sqrt{\frac{\hbar}{m_\perp\omega_r}} = c_\perp \sqrt{\frac{2}{\omega_r \gamma H}},\qquad
a_z=\sqrt{\frac{\hbar}{m_\parallel\omega_z}} = c_\parallel \sqrt{\frac{2}{\omega_z \gamma H}}.
\end{equation}
Usually we are interested only in non-antisymmetric states which
have non-zero integral transverse magnetization and can be excited and
observed in NMR experiments. For these states $n_\phi = 0$, $n_z =
0,2,4\ldots$, and
\begin{equation}\label{eq1:wave_form}
s_+(r,z) =
\ \left(\pi^{3/2} a_r^2 a_z \ 2^{n_z}\ n_z!\right)^{-1/2}
\ \exp\left(-\frac{r^2}{2a_r^2}-\frac{z^2}{2a_z^2}\right)
L_{n_r}^0\left(\frac{r^2}{a_r^2}\right)
H_{n_z}\left(\frac{z}{a_z}\right),
\end{equation}
\begin{equation}
\omega = \omega_0 + \omega_r(2n_r+1) + \omega_z(n_z+1/2),\qquad
n_r=0,1,2\ldots, n_z=0,2,4\ldots
\end{equation}

\eject
\section*{Part 2. Non-linear spin dynamics}

\subsection*{Leggett equations in {}$R_{aj}$ coordinates}

In this part non-linear Leggett equations of spin dynamics in
$^3$He-B~\cite{1975_he3_teor_leggett} are derived. We start from same
Hamiltonian as in the first part and use components of matrix~$R_{aj}$ as
coordinates.

Order parameter of $^3$He-B is
\begin{equation}
A_{aj}  = \frac{1}{\sqrt{3}}\ \Delta\ e^{i\varphi} R_{aj},
\end{equation}
where~$\varphi$ is the phase, and $R_{aj}$
is a rotation matrix with axis~$\bf n$ and rotation angle~$\vartheta$:
\begin{equation}\label{eq:r_nt}
R^0_{a j} = \ct\ \delta_{a j} + (1-\ct)\ n_a n_j - \st\ e_{ajk} n_k.
\end{equation}
Hamiltonian of this system is:
\begin{equation}\label{eq:ham}
\mathcal{H} = F_M + F_{SO} + F_\nabla,
\end{equation}
\begin{eqnarray}
\label{eq:en_m0}
F_M &=& - ({\bf S} \cdot \gamma {\bf H})
+ \frac{\gamma^2}{2\chi_B}{\bf S}^2,\\
\label{eq:en_d0}
F_{SO}
&=& g_D\Delta^2 \Big[
           R_{jj}R_{kk}
         + R_{jk}R_{kj}
 - \frac23 R_{jk}R_{jk}\Big]
= g_D\Delta^2 \Big[ R_{jj}R_{kk} + R_{jk}R_{kj}] + \mbox{const.}
,\\
\label{eq:en_g0}
F_\nabla
&=& \frac12 \Delta^2 \Big[
  K_1 (\nabla_j R_{ak})(\nabla_j R_{ak})
+ K_2 (\nabla_j R_{ak})(\nabla_k R_{aj})
+ K_3 (\nabla_j R_{aj})(\nabla_k R_{ak}) \Big].
\end{eqnarray}
Equations of motion are
\begin{eqnarray}\label{eq:shs}
\dot S_a
&=& \{\mathcal{H}, S_a\}
\quad=\quad
  \frac{\delta \mathcal{H}}{\delta S_b} \{S_b,S_a\}
+ \frac{\delta \mathcal{H}}{\delta R_{bj}} \{R_{bj},S_a\}
\quad=\quad
  \frac{\delta \mathcal{H}}{\delta S_b} e_{abc} S_c
+ \frac{\delta \mathcal{H}}{\delta R_{bj}} e_{abc} R_{cj},
\\\label{eq:shr}
\dot R_{aj}
&=& \{\mathcal{H}, R_{aj}\}
\quad=\quad
  \frac{\delta \mathcal{H}}{\delta S_b} \{S_b,R_{aj}\}
+ \frac{\delta \mathcal{H}}{\delta R_{bk}} \{R_{bk},R_{aj}\}
\quad=\quad
  \frac{\delta \mathcal{H}}{\delta S_b} e_{abc} R_{cj}.
\end{eqnarray}
The derivatives of the Hamiltonian are much simpler then in the case
of~$\ts$ angles. There are no commutation problem in the variational
derivative. The result is:
\begin{eqnarray}
\frac{\delta \mathcal{H}}{\delta {\bf S}} &=&
- \gamma {\bf H} + \frac{\gamma^2}{\chi_B}{\bf S},
\\
\frac{\delta \mathcal{H}}{\delta R_{bj}}
 &=& \Delta^2 \big[
  2 g_D(\delta_{bj}R_{kk} + R_{jb} - 2/3 R_{bj})
- K_1 (\nabla_k \nabla_k R_{bj})
- (K_2+K_3) (\nabla_k \nabla_j R_{bk})
\big].
\end{eqnarray}
Substituting this into equations of motion~(\ref{eq:shs},\ref{eq:shr}) we obtain
Leggett equations:
\begin{eqnarray}\label{eq:leggett_r}
\dot S_a &=&
  [{\bf S}\times \gamma {\bf H}]_a + T^D_a + T^\nabla_a,
\\\nonumber
\dot R_{aj} &=&
  e_{abc} R_{cj} \Big(\frac{\gamma^2}{\chi_B} {\bf S} - \gamma {\bf H} \Big)_b,
\end{eqnarray}
with dipolar and gradient torques:
\begin{eqnarray}\label{eq:torqueR}
T^D_a
 &=& e_{abc} \frac{\delta F_D}{\delta R_{bj}} R_{cj}
 \quad=\quad
  2 \Delta^2 g_D\ e_{abc} R_{cj} (\delta_{bj}R_{kk} + R_{jb}),
\\\nonumber
T^\nabla_a
 &=& e_{abc} \frac{\delta F_\nabla}{\delta R_{bj}} R_{cj}
 \quad=\quad
 - \Delta^2 e_{abc} R_{cj} \big[
     K_1 (\nabla_k \nabla_k R_{bj})
    + (K_2+K_3) (\nabla_k \nabla_j R_{bk}) \big].
\end{eqnarray}
The gradient torque can be rewritten in terms of spin current~$J_{ak}$ which
carries $a$ component of the spin in the direction $k$:
\begin{equation}\label{eq:spin_curr}
T^\nabla_a = - \nabla_k J_{ak},
\qquad
J_{ak} = \Delta^2  e_{abc} R_{cj} \big[ K_1 (\nabla_k R_{bj}) + (K_2+K_3) (\nabla_j R_{bk}) \big].
\end{equation}
Here we used equality~$e_{abc}\nabla_k R_{bj}\nabla_k R_{cj} =
e_{abc}\nabla_k R_{bj}\nabla_j R_{ck} = 0$.

It is also interesting to study motion of orbital anisotropy
vector~$L_j = S_aR_{aj}$:
\begin{equation}
\dot L_k
\quad=\quad
\{\mathcal{H},  S_aR_{ak}\}
\quad=\quad
\{\mathcal{H}, S_a\} R_{ak} +
\{\mathcal{H}, R_{ak}\} S_a 
\quad=\quad
\dot S_a R_{aj} + \dot R_{aj} S_a.
\end{equation}
Substituting equations for $\dot S_a$ and $\dot R_{aj}$ we have:
\begin{eqnarray}
\dot L_k &=& (T^D_a + T^\nabla_a) R_{ak}.
\end{eqnarray}

\subsection*{Leggett equations in ${\bf n}$ and $\vartheta$ coordinates} 

Let's rewrite Leggett equations in ${\bf n}$ and $\vartheta$ coordinates
(axis and angle or rotation of the matrix $R_{aj}$).

Time derivative of matrix~$R({\bf n}, \vartheta)$ is:
\begin{equation}
\dot R_{aj} = \big[ \st\ (n_a n_j - \delta_{aj}) - \ct\ e_{ajk} n_k \big]\,\dot \vartheta
 + (1-\ct) [ n_j \dot n_a + \ n_a \dot n_j ] - \st\ e_{ajk} \dot n_k.
\end{equation}
Let's calculate following combinations:
\begin{eqnarray}
n_a \dot R_{aj} + n_a \dot R_{ja} &=& 2 (1-\ct)\ \dot n_j,
\\\nonumber
R_{ja} \dot R_{aj} &=& -4\ct\st\ \dot\vartheta.
\end{eqnarray}
By substituting $R$ from~(\ref{eq:r_nt}) and
$\dot R$ from~(\ref{eq:leggett_r}) we can write Leggett equations in $\dot n$ and $\dot\theta$ coordinates:
\begin{eqnarray}\label{eq:leggett_nt}
\dot {\bf S} &=& {\bf S}\times \gamma {\bf H}  + {\bf T}^D - \nabla_k J_{ak},
\\\nonumber
{\bf\dot n} &=&
-\frac12\ {\bf n}\times \Big[
\Big(\frac{\gamma^2}{\chi_B} {\bf S} - \gamma {\bf H} \Big)
+ \frac{\st}{1-\ct}
\ {\bf n}\times \Big(\frac{\gamma^2}{\chi_B} {\bf S} - \gamma {\bf H} \Big)
\Big],
\\\nonumber
\dot\vartheta &=&  {\bf n} \cdot \Big(\frac{\gamma^2}{\chi_B} {\bf S} - \gamma {\bf H} \Big),
\end{eqnarray}
Expression for dipolar torque via ${\bf n}$ and
$\vartheta$:
\begin{equation}
{\bf T}^D = 4 \Delta^2 g_D\ \st(4\ct + 1)\ {\bf n}.
\end{equation}
%
%
%
Spin current:
\begin{eqnarray}
J_{ak} &=&  2\Delta^2 K1 \big[
  - n_a \nabla_k\vartheta
  - \st\ \nabla_k n_a
  + (1-\ct)\ e_{abc} n_c \nabla_k n_b
\big]\\\nonumber
&+& \Delta^2 (K_2+K_3) \big[
 -             n_a             \nabla_k\vartheta
 + \ct\        \delta_{ak} n_j \nabla_j\vartheta
 + (1-\ct)\    n_k n_j n_a     \nabla_j\vartheta
\\\nonumber
&&
+ \ct\st\     e_{abj} n_b n_k \nabla_j\vartheta
+ (1-\ct)\st\ e_{abk} n_b n_j \nabla_j\vartheta
\\\nonumber
&&
 - \ct\st\     e_{kbj} n_b n_a \nabla_j\vartheta
 - \ct\st\     e_{akj}         \nabla_j\vartheta
\\\nonumber
&&
 + \st\ct         \delta_{ak} \nabla_j n_j
 + (1-\ct)\st     n_a n_k     \nabla_j n_j
 + (1-\ct)\st     n_a n_j     \nabla_j n_k
\\\nonumber
&&
 - (1-\ct)\st     n_k n_j     \nabla_j n_a
 - \sts           \delta_{ak} e_{cjb} n_b \nabla_j n_c
 + (1-\ct)^2      e_{abc} n_c n_j n_k \nabla_j n_b
\\\nonumber
&&
 + (1-\ct)\ct     e_{abj} n_k \nabla_j n_b
 + (1-\ct)\ct     e_{abj} n_b \nabla_j n_k
 - \ct\st                     \nabla_k n_a
\\\nonumber
&&
 - (1-\ct)\st                 \nabla_a n_k
 + \sts           e_{kjb} n_b  \nabla_a n_j
\big].
\end{eqnarray}
Expression for gradient torque can be obtain by finding spin current derivative~(\ref{eq:spin_curr}).

\subsection*{
Spin created by motion of ${\bf n}$ and $\vartheta$
}

It could be useful to solve second and third Leggett equations~(\ref{eq:leggett_nt}) for spin. We use
temporary notation
\begin{equation}
{\bf s} = \frac{\gamma^2}{\chi_B} {\bf S} - \gamma
{\bf H},
\qquad
C = \frac{\st}{1-\ct}.
\end{equation}
Then the equations are:
\begin{equation}
{\bf\dot n} =
-\frac12\ {\bf n}\times \Big[
{\bf s}
+ C\ {\bf n}\times {\bf s}
\Big],
\qquad
\dot\vartheta =  {\bf n} \cdot {\bf s}
\end{equation}
Let's solve them for {\bf s}. Solution can be written as ${\bf s} = {\bf
n}\dot\vartheta + {\bf n}\times{\bf A}$ with ${\bf A}\cdot{\bf n} = 0$.
This satisfies the equation for~$\dot\vartheta$ and splits vector~$\bf s$ into two
parts, parallel and perpendicular to~$\bf n$.
Substitute~$\bf s$ into the equation for~$\bf\dot n$:
\begin{equation}
{\bf A} + C {\bf n}\times{\bf A} = 2 {\bf\dot n},
\end{equation}
find vector product of this equation with $\bf n$, multiply it by $C$ and
subtract from original equation. From this we can find
${\bf A} = (1-\ct)\ {\bf\dot n} - \st\ {\bf n}\times{\bf\dot n}$
and get the result:
\begin{equation}\label{eq:S_dndt}
\frac{\gamma^2}{\chi_B} {\bf S} =
\gamma {\bf H} + {\bf n}\ \dot\vartheta + (1-\ct)\ {\bf n}\times{\bf\dot n} + \st\ {\bf\dot n}.
\end{equation}
Here one can see how spin is created by magnetic field and motion of ${\bf n}$ and
$\vartheta$.

There is an interesting observation: if we have a soliton moving with a
constant velocity, we can have a uniform spin distribution in the moving
frame, but non-uniform in the static one.

\subsection*{
Leggett equations for ${\bf n}$ and $\vartheta$ in the rotating frame
}

For numerical integration of Leggett equations~(\ref{eq:leggett_nt}) it is
convenient to switch to a ``rotating frame'' by doing substitution
\begin{equation}
S^r_{a'} = U_{a'a} S_a,\quad
T^r_{a'} = U_{a'a} T_a,\quad
R^r_{a'k'} = U_{k'k} U_{a'a} R_{ak},\quad\ldots
\end{equation}
with rotation matrix
$U_{jk} = R_{jk}(\omega t, {\bf\hat z}) = \cos\omega t\ \delta_{j k} + (1-\cos\omega t)\ {\bf\hat z}_j {\bf\hat z}_k - \sin\omega t\ e_{jkl}{\bf\hat z}_l$.

For me it is still not clear is this ``rotating frame'' has a physical
meaning (for example in 3D space we do not specify any axis of rotation).
Anyway, we can always do the substitution, solve equations and return
back to original vectors if needed.

By making the substitution in~(\ref{eq:leggett_r}) and using facts that
$U{\bf a}\cdot U{\bf b} = {\bf a\cdot b}$, $[U{\bf a}\times U{\bf b}] = U
[{\bf a}\times {\bf b}]$ and $U \dot U^{-1} {\bf a} = \omega [{\bf a}
\times {\bf\hat z}]$ we have equations in the rotating frame:
\begin{eqnarray}\label{eq:r_motion_rot}
 \dot {\bf S}^r + \omega [{\bf S}^r \times {\bf\hat z}]  &=&
  [{\bf S}^r\times \gamma {\bf H}^r] + {\bf T}^{Dr} + {\bf T}^{\nabla r}
\\\nonumber
{\bf\dot n}^r + \omega [{\bf n}^r \times {\bf\hat z}] &=&
-\frac12\ {\bf n}^r\times \Big [ \Big(\frac{\gamma^2}{\chi_B} {\bf S}^r - \gamma {\bf H}^r \Big)
+ \frac{\st}{1-\ct}\ {\bf n}^r\times \Big(\frac{\gamma^2}{\chi_B} {\bf S}^r - \gamma {\bf H}^r \Big)
\Big]
\\\nonumber
\dot\vartheta &=& {\bf n}^r \cdot \Big(\frac{\gamma^2}{\chi_B} {\bf S}^r - \gamma {\bf H}^r \Big).
\end{eqnarray}
with dipolar torque
\begin{eqnarray}\label{eq:td_nt}
{\bf T}^{Dr} = 4 \Delta^2 g_D\ \st(4\ct + 1)\ {\bf n}^r
\end{eqnarray}
Writing the gradient torque in the rotating frame is not trivial, because
matrix $U$ will act on gradient operators. To avoid this difficulty we
will be solving only 1D problems with all gradients directed along
${\bf\hat z}$ axis. Then spin current is not affected by rotation and we
can simply write:
\begin{equation}
{\bf T}^{\nabla r} = -\nabla_3 J_{a3} = -{\bf J}',
\qquad
J_{a} =
 \Delta^2  e_{abc} R_{cj} \big[ K_1 R_{bj}' + (K_2+K_3) R_{b3}'\delta_{j3} \big],
\end{equation}
using primes for $\nabla_3$. Usually we assume that magnetic field in the rotating frame~${\bf H^r}$
is constant and choose axis~$\hat {\bf x}$ in such a way
that field is in $x-z$ plane:
\begin{equation}
\label{eq:def_first}
  \gamma{\bf H^r} = {\bf\hat z}\,\omega_z + \hat {\bf x}\,\omega_x
\end{equation}
This is a good description of a usual continuous-wave NMR experiments with
static field $\gamma H_z = \omega_z$ and rotating radio-frequency
field $\gamma H_x = \omega_x$.

\subsection*{Spin in the rotating frame}

In the following calculations we will work only with vectors in the
rotating frame and omit indices~``$r$''.
It's convenient to introduce a dimensionless vector~$\bf w$:
\begin{equation}\label{eq:w}
{\bf w}
\ =\ %
\frac{1}{\omega}\Big(\frac{\gamma^2}{\chi_B} {\bf S} - \gamma {\bf H} \Big) + {\bf\hat z}
\end{equation}
Then Leggett equations will be written as
\begin{eqnarray}\label{eq:r_motion_wnt_eq}
  \dot{\bf w}  &=&
  {\bf w}\times ({\bf\hat z}\,(\omega_z-\omega) + \hat {\bf x}\,\omega_x)
 + \frac{\gamma^2}{\omega\chi_B}\left({\bf T}^{D} + {\bf T}^{\nabla}\right)
\\\nonumber
{\bf\dot n} / \omega &=&
-\frac12\ {\bf n}\times \Big[
( {\bf w} + {\bf\hat z})
+ \frac{\st}{1-\ct}
\ {\bf n}\times ({\bf w} - {\bf\hat z})
\Big]
\\\nonumber
\dot\vartheta / \omega &=& {\bf n} \cdot ({\bf w} - {\bf\hat  z}).
\end{eqnarray}
We can solve second and third equations for~$\bf w$ in the same way as~(\ref{eq:S_dndt}):
\begin{eqnarray}
{\bf w} &=&
  \ct\ {\bf\hat z}
+ (1-\ct)\ {\bf n}\ n_z
+ \st\ {\bf n}\times {\bf\hat z}
\\\nonumber
&+&
  \frac1{\omega}\Big[ {\bf n}\ \dot\vartheta
+ (1-\ct)\ {\bf n}\times{\bf\dot n}
+ \st\ {\bf\dot n}
\Big]
\end{eqnarray}
This can be also written as:
\begin{equation}\label{eq:wr_dndt}
{\bf w} = R{\bf\hat z}
+ \frac1{\omega}\left[ {\bf n}\ \dot\vartheta
+ \frac{1-\ct}{\st}\ (R{\bf\dot n} + {\bf\dot n})
\right]
\end{equation}
Or in original notation:
\begin{equation}
\frac{\gamma^2}{\chi_B} {\bf S}
=
\gamma {\bf H}
+ {\bf n}\ \dot\vartheta
+ (1-\ct)\ {\bf n}\times[{\bf\dot n} + \omega {\bf n}\times{\bf\hat z}]
+ \st\ [{\bf\dot n} + \omega {\bf n}\times {\bf\hat z}]
\end{equation}
\begin{equation}\label{eq:Sr_dndt}
\frac{\gamma^2}{\chi_B} {\bf S}
=
\gamma {\bf H}
+ \omega\ (R{\bf\hat z}-{\bf\hat z})
+ {\bf n}\ \dot\vartheta
+ \frac{1-\ct}{\st}\ (R{\bf\dot n} + {\bf\dot n})
\end{equation}

\subsection*{Non-uniform equilibrium in the rotating frame}

Equilibrium state can be found by setting all time derivatives to zero.
From~(\ref{eq:wr_dndt}) we have
\begin{equation}\label{eq:w_n}
w_a = R_{a3},
\end{equation}
it means that in the equilibrium $|{\bf w}| = 1$.

Using this equation we can exclude $\bf w$ from the first equation and
write equilibrium problem using only $R_{aj}$ (or $\bf n$ and $\vartheta$) coordinates:
\begin{equation}\label{eq:hpd_texture1}
e_{abc} R_{bj} \Big[
\frac{\omega\chi_B}{\gamma^2 \Delta^2}\left(\delta_{c3}\,(\omega_z-\omega) + \delta_{c1}\,\omega_x\right) \delta_{j3}
-  2g_D
\ (\delta_{cj}R_{kk} + R_{jc})
+ K_1 R_{cj}'' + (K_2+K_3)\delta_{j3} R_{c3}''
\Big] = 0.
\end{equation}
Here we use expressions~(\ref{eq:torqueR}) for dipolar and gradient
torques and keep only gradients in $\bf\hat z$ direction, represented by
primes.

This can be written in term of Leggett frequency and spin-wave velocities:
\begin{equation}\label{eq:hpd_texture2}
e_{abc} R_{bj} \Big[
\omega(\left(\omega_z-\omega)\ \delta_{c3} + \omega_x\ \delta_{c1}\right)\delta_{j3}
-  \frac{2}{15}\Omega_B^2 \ (\delta_{cj}R_{kk} + R_{jc})
+ \frac12 C_{\parallel}^2 R_{cj}''
+ (C_{\perp}^2 - C_{\parallel}^2)\delta_{j3} R_{c3}''
\Big] = 0.
\end{equation}
This is a set of three equations for three parameters~($\bf n$ and $\vartheta$). It
describes non-uniform texture of a precessing state and
can be used for calculating HPD boundaries or solitons.

\subsection*{Uniform equilibrium}

Let's remove gradient terms in the first equation
in~(\ref{eq:r_motion_wnt_eq}) and use expression~(\ref{eq:td_nt}) for
dipolar torque:
\begin{equation}\label{eq:eq_wn}
{\bf w}\times({\bf\hat z}\,(\omega_z-\omega) + {\bf\hat x}\,\omega_x)
= \frac{4}{15}\frac{\Omega_B^2}{\omega}\ \st(4\ct + 1)\ {\bf n},
\end{equation}
By multiplying it by $\bf w$ we have left-hand side equals zero and three cases:
${\bf n}\cdot{\bf w} = 0$, or $\ct = -1/4$, or $\st=0$.
The last one is unstable equilibrium, it corresponds to maximum of
dipolar energy. Two other cases are called HPD and NPD, now we can
find them.

\medskip
{\bf HPD (Homogeneously-precessing domain):}

We already showed in~(\ref{eq:w_n}) that
${\bf w} = R {\bf\hat z}$. Using this,~(\ref{eq:eq_wn}), and HPD condition ${\bf n}\cdot{\bf w} = 0$
we can find
\begin{equation}\label{eq:hpd_eq1}
{\bf n} = \pm{\bf\hat y},
\qquad
{\bf w} = \ct\ {\bf\hat z} +  n_y\st\ {\bf\hat x}.
\end{equation}
\begin{equation}\label{eq:hpd_eq2}
\ct =
- \frac14 + \frac{15}{16}\frac{\omega}{\Omega_B^2}
\ \left((\omega-\omega_z) + n_y \frac{\ct}{\st}\,\omega_x\right).
\end{equation}
We can use only positive sign of $\vartheta$ because matrix $R({\bf n},
\vartheta)$ is same as $R(-{\bf n}, -\vartheta)$. Later it will be shown
that state with $\vartheta>0$ and ${\bf n} = -{\bf\hat y}$ is unstable.
To check it we will keep parameter $n_y = \pm1$, which can be always set to 1
for HPD equilibrium state.

Usually RF-pumping $\omega_x$ is much smaller then frequency shift $\omega-\omega_z$,
$\ct$ is close to -1/4 and we can use $\ct/\st = -1/\sqrt{15}$ in the right part
of the last equation:
\begin{equation}\label{eq:hpd_eq2a}
\ct =
- \frac14 + \frac{15}{16}\frac{\omega}{\Omega_B^2}
\ \left(\omega-\omega_z - \frac{n_y}{\sqrt{15}}\,\omega_x\right).
\end{equation}
In arbitrary case to find $\vartheta$ we can do substitutions
\begin{equation}
t = \tan\vartheta/2,
\qquad
A = \frac{15}{16}\frac{\omega(\omega-\omega_z)}{\Omega_B^2},
\qquad
B = \frac{15}{16}\frac{\omega}{\Omega_B^2} n_y\,\omega_x,
\end{equation}
and obtaining 4-th order equation for $t$:
\begin{equation}
Bt^4 - \frac12(3 + 4A)t^3 + \frac12(5 - 4A)t - B = 0.
\end{equation}

\medskip
{\bf NPD (Non-precessing domain), Brinkman-Smith mode:}

Using the NPD condition $\ct = -1/4$ we have:
\begin{equation}
w_1 = \frac{-\omega_x}{\sqrt{\omega_x^2 + (\omega_z-\omega)^2}},
\qquad
w_2 = 0,
\qquad
w_3 = \frac{-(\omega_z-\omega)}{\sqrt{\omega_x^2 + (\omega_z-\omega)^2}},
\end{equation}
\begin{equation}
n_1 =  \frac{w_1 n_3}{1+w_3},
\qquad
n_2 = \ \frac{w_1\sqrt{3/5}}{1+w_3},
\qquad
n_3 = \pm(4w_3+1)/5.
\end{equation}
(There is also solution with opposite signs of $w_1$ and $w_3$ which
corresponds to unstable equilibrium).

Here $\bf n$ is precessing around~$\bf\hat z$ axis. This is
case of a ``vertical'' texture. In real NMR experiments if RF pumping is
small enough then texture is determined by boundary conditions in the
experimental cell and possible topological defects such as $\bf
n$-solitons. It can have any equilibrium orientation of~$\bf n$ vector
and small oscillations around this equilibrium. These solutions can not
be found as an equilibrium in our rotating frame.


\subsection*{Small oscillations of HPD in the rotating frame}

This section follows calculations in my PhD thesis (2012, available only
in Russian).

Let's again consider a uniform case with zero gradient terms. Introduce
dimensionless parameters:
\begin{equation}\label{eq:defs1}
d = \frac{\omega-\omega_z}{\omega},
\qquad
h = \frac{\omega_x}{\omega},
\qquad
b = \frac{15\gamma^2\Delta^2 g_D}{\chi_B\ \omega^2} = \frac{\Omega_B^2}{\omega^2}
\end{equation}

Then equations~(\ref{eq:r_motion_wnt_eq}) will be
\begin{eqnarray}\label{eq:r_motion_wnt}
\dot {\bf w}/\omega  &=&
 {\bf w} \times (h\,\hat {\bf x} - d\,{\bf\hat z})
  + \frac{4b}{15}\ \st(4\ct + 1)\ {\bf n}
\\\nonumber
{\bf\dot n}/\omega &=&
-\frac12\ {\bf n}\times ({\bf w} + {\bf\hat z})
-\frac12\ \frac{\st}{1-\ct}\ \Big[ {\bf n}\ ({\bf n}\cdot({\bf w} - {\bf\hat z}))
- ({\bf w} - {\bf\hat z})\Big],
\\\nonumber
\dot\vartheta/\omega &=& {\bf n} \cdot ({\bf w} - {\bf\hat z}),
\end{eqnarray}
(In my PhD thesis slightly different definition of dimensionless parameters were
used. Also there was an error in equilibrium magnetization values, but it
does not change the answer for small oscillation frequencies.)

The equilibrium which corresponds to HPD~(\ref{eq:hpd_eq1},\ref{eq:hpd_eq2a}) is
\begin{equation}
{\bf n}^0 = \pm{\bf\hat y},
\qquad
{\bf w}^0 = \ct\ {\bf\hat z} +  n_y\st\ {\bf\hat x}
\qquad
\ct^0 = -\frac{1}{4} - \frac{15}{16} \Big(d - \frac{h\,n_y}{\sqrt{15}}\Big)\frac{1}{b}
\end{equation}

Let's rewrite equation~(\ref{eq:r_motion_wnt}) in coordinates:
\begin{eqnarray}
\dot w_x/\omega &=&
F\,b\ n_x - w_y d
\\\nonumber
\dot w_y/\omega &=&
F\,b\ n_y + w_z h + w_x d
\\\nonumber
\dot w_z/\omega &=&
F\,b\ n_z - w_y h
\\\nonumber
2 \dot n_x/\omega &=&
- n_y (w_z + 1) + n_z w_y
-C\ n_x\ ({\bf n}\cdot({\bf w} - {\bf\hat z}))
+ C\ w_x
\\\nonumber
2 \dot n_y/\omega &=& - n_z w_x + n_x (w_z + \omega)
-C\ n_y\ ({\bf n}\cdot({\bf w} - {\bf\hat z}))
+ C\ w_y
\\\nonumber
2 \dot n_z/\omega &=& - n_x w_y + n_y w_x
- C\ n_z\ ({\bf n}\cdot({\bf w} - {\bf\hat z}))
+ C\ (w_z - 1)
\\\nonumber
\dot\vartheta/\omega &=& {\bf n} \cdot ({\bf w} - {\bf\hat z}).
\end{eqnarray}
where
\begin{equation}\label{eq:CF}
C=\frac{\st}{1-\ct},\qquad
F=\frac{4}{15}\ \st(4\ct + 1).
\end{equation}

Three equations for $\dot{\bf n}$, are dependent because $|{\bf n}|=1$. This
can  be checked by calculating~$n_x\dot n_x + n_y\dot n_y + n_z\dot n_z$.

Lets write the equations for small deviations $d{\bf n}$, $d{\bf w}$, $d\vartheta$
from the equilibrium. Here we keep only first-order terms with the deviations.
Also we put $dn_y = 0$ and omit equation for~$n_y$ because $\bf n$ is moving near $\hat{\bf y}$:
\begin{eqnarray}
d \dot w_x/\omega &=& F^0\,b\ dn_x - d\,dw_y,
\\\nonumber
d \dot w_y/\omega &=& dF\,b\ n^0_y + h\,dw_z + d\,dw_x,
\\\nonumber
d \dot w_z/\omega &=& F^0\,b\ dn_z - h\,dw_y,
\\\nonumber
2 d\dot n_x/\omega &=& - n^0_y\ dw_z + C^0\ dw_x + w^0_x\ dC,
\\\nonumber
2 d\dot n_z/\omega &=&   n^0_y\ dw_x + C^0\ dw_z + (w_z-1)\ dC,
\\\nonumber
d\dot\vartheta/\omega &=& w^0_x\ dn_x  + (w^0_z-1)\ dn_z + n^0_y\ dw_y.
\end{eqnarray}
For periodic motion with dimensionless frequency~$x = \Omega/\omega$ this
gives an equation
\begin{equation}
\left|
\begin{array}{rrrrrr}
-ix &  -d & 0   & b\,F^0 & 0 & 0\\
 d  & -ix & h   & 0 & 0 & n^0_y b\,dF/d\vartheta\\
0   & -h  & -ix & 0 & b\,F^0 & 0\\
C   &  0  & -n_y & -2ix & 0    & n^0_y \st\ dC/d\vartheta\\
n_y &  0  &    C & 0    & -2ix & (\ct^0 - 1)\ dC/d\vartheta\\
    0&  n^0_y&  0& w^0_x & (\ct^0 - 1) & -ix
\end{array}
\right| = 0
\end{equation}

If~$d,h \ll b \ll 1$, then the equation is
\begin{equation}
-x^6 + (1 + b)\,x^4 - \frac{1}{\sqrt{15}}
\left(\frac38 (\sqrt{15}d - n_y h) + b (\sqrt{15}d + 3 n_y h)\right)\,x^2
+ \frac{1}{10}(\sqrt{15}d -n_y h)\,b\,n_y h = 0;
\end{equation}
where~$n^0_y=\pm1$ and~$\vartheta^0$ is positive (changing sign of~$\vartheta$ is
equivalent to changing sign of~$\bf n$).

Solutions are
\begin{eqnarray}
x_1^2 &=& 1+b, \\\nonumber
x_2^2 &=& \frac{4}{\sqrt{15}}\ \frac{n^0_y h\,b}{1+8/3\,b}, \\\nonumber
x_3^2 &=& \frac{\sqrt{15}\,d - n^0_y h}{\sqrt{15}}
          \ \frac{3/8 + b}{1+b}.
\end{eqnarray}
One can see that $n^0_y = -1$ corresponds to unstable state ($x_2^2 < 0$) and we can use $n^0_y=+1$.

In original notations
\begin{eqnarray}
\Omega_1^2 &=& \omega^2 + \Omega_B^2, \\\nonumber
\Omega_2^2 &=& \frac{4}{\sqrt{15}}\ \frac{\omega_x\omega\,\Omega_B^2}{\omega^2+8/3\,\Omega_B^2}, \\\nonumber
\Omega_3^2 &=& \frac{\sqrt{15}\,(\omega-\omega_z)\omega - \omega_x\omega}{\sqrt{15}}
          \ \frac{3/8\,\omega^2 + \Omega_B^2}{\omega^2 + \Omega_B^2}.
\end{eqnarray}
The second mode is a useful tool for calibrating RF pumping~$\omega_x$ or
measuring Leggett frequency~$\Omega_B$. The third mode as fas as I know
have not been observed in experiments.

\subsection*{Leggett-Takagi relaxation} 

Leggett-Takagi equations~\cite{1977_leggett_takagi} can be made from Leggett equations~(\ref{eq:leggett_nt})
by adding a relaxation term with parameter~$\tau$ to the equation for~$\dot\vartheta$.
\begin{eqnarray}
\dot\vartheta &=&  {\bf n} \cdot \Big(\frac{\gamma^2}{\chi_B} {\bf S} - \gamma {\bf H} \Big)
+ \frac4{15}\ \st(4\ct + 1)\ \frac1{\tau}.
\end{eqnarray}

Switching to the rotating frame and using dimensionless parameters ${\bf
w}, d, h, b$~(\ref{eq:w}, \ref{eq:defs1}) we can write equations of
motion~(\ref{eq:r_motion_wnt}) with the relaxation term:
\begin{eqnarray}\label{eq:r_motion_lt}
\dot {\bf w}/\omega  &=&
 {\bf w} \times (-d\,\hat {\bf z} + h\,\hat {\bf x})
  + \frac{4}{15}b\ \st(4\ct + 1)\ {\bf n}
\\\nonumber
\dot {\bf n}/\omega &=&
-\frac12\ {\bf n}\times ({\bf w} + \hat {\bf z})
-\frac12\ \frac{\st}{1-\ct}\ \Big[ {\bf n}\ ({\bf n}\cdot({\bf w} - \hat {\bf z}))
- ({\bf w} - \hat {\bf z})\Big]
\\\nonumber
\dot\vartheta/\omega &=& {\bf n} \cdot ({\bf w} - \hat {\bf z})
+ \frac4{15}\ \st(4\ct + 1)\ \frac1{\omega\tau},
\end{eqnarray}
or in coordinates with $C$ and $F$ defined in~(\ref{eq:CF}):
\begin{eqnarray}\label{eq:r_motion_lt}
\dot w_x/\omega  &=& -d\ w_y + bF\ n_x \\\nonumber
\dot w_y/\omega  &=&  h\ w_z + d\ w_x + bF\ n_y \\\nonumber
\dot w_z/\omega  &=& -h\ w_y  + bF\ n_z\\\nonumber
2\dot n_x/\omega &=& -n_y (w_z + 1) + n_z w_y
-C\ n_x\ ({\bf n}\cdot({\bf w} - \hat {\bf z})) + C w_x
\\\nonumber
2\dot n_y/\omega &=& -n_z w_x + n_x (w_z + 1)
-C\ n_y\ ({\bf n}\cdot({\bf w} - \hat {\bf z})) + C w_y
\\\nonumber
2\dot n_z/\omega &=& -n_x w_y +n_y w_x
-C\ n_z\ ({\bf n}\cdot({\bf w} - \hat {\bf z})) + C (w_z - 1)
\\\nonumber
\dot\vartheta/\omega &=& {\bf n} \cdot ({\bf w} - \hat {\bf z})
+ \frac{F}{\omega\tau}.
\end{eqnarray}

For the equilibrium state we have:
\begin{equation}
{\bf n}^0\cdot{\bf w}^0 = 0,
\qquad
n^0_z = \frac{F^0}{\omega\tau},
\qquad
n^0_x = \frac{d}{h}\ \frac{F^0}{\omega\tau},
\qquad
w^0_y  = \frac{b}{h}\ \frac{(F^0)^2}{\omega\tau}.
\end{equation}
One can see that equilibrium position of vector $\bf n$ is rotated from
${\bf\hat y}$ direction by Leggett-Takagi relaxation, and vector~${\bf
w}^0$ stays perpendicular to ${\bf n}^0$. Note that even at low
relaxation~$n^0_x$ can be large because of~$d/h$ factor. This probably
determines maximum frequency shift where HPD can exist for a given RF
pumping and Leggett-Takagi relaxation. It is very interesting to measure
this effect in experiment.

\subsection*{$\vartheta$-solitons} 

Minimum of spin-orbit interaction corresponds to $\ct = -1/4$. It is
possible to have a so-called theta-soliton~\cite{1976_maki_solitons}
there $\vartheta$ changes between two minima, $\vartheta_L =
\cos^{-1}(-1/4)$ and $2\pi-\vartheta_L$, as a solution of equilibrium
Leggett equations.

Let's assume ${\bf n} = \hat{\bf z}$, ${\bf H} = H_z\hat{\bf z}$, and ${\bf
S} = \chi_B {\bf H}/\gamma$. Only $\vartheta$ is changing along $\hat{\bf
z}$ axis.

Using usual expression $R_{a j} = \ct\ \delta_{a j} + (1-\ct)\ n_a n_j - \st\ e_{ajk} n_k$,
and $n_j = \delta_{j3}$ one can find that only $J_{33}$ component of spin current is non-zero:
\begin{eqnarray}
J_{33} &=& - 2 \Delta^2 K_1 \vartheta'
\end{eqnarray}
Second and third of Leggett equations are satisfied by chosen
values of ${\bf n}$ and ${\bf S}$. The first equation gives
non-trivial expression for distribution of $\vartheta$:
\begin{equation}
T^D_3 = J_{33}'
\end{equation}
By substituting spin current and dipolar torque we obtain a second-order
differential equation for $\vartheta$:
\begin{equation}\label{eq:th_sol0}
\vartheta'' = -\frac{2 g_D}{K1}\ \st(4\ct + 1)
\end{equation}
or with usual definition of dipolar length~$\xi_D^2 = \frac{13}{24}\frac{K_1}{g_D}$.
\begin{equation}
\frac{12}{13}\xi_D^2\ \vartheta'' = -\st(4\ct + 1)
\end{equation}

To solve this equation we multiply both sides by $\vartheta'$ and integrate using
boundary conditions at $z=\pm\infty$: $\cos\vartheta = -1/4$, $\vartheta'=0$:
\begin{equation}
\sqrt{\frac{12}{13}} \xi_D\ \vartheta' = \pm 2 (\cos\vartheta + 1/4)
\end{equation}

Second integration will give us formula of the $\vartheta$-soliton.
Integration constant determines position of the soliton, we choose it to
localize the soliton at $z=0$. There are two solutions, one goes from
$-\theta_L$ to $+\theta_L$, another from $+\theta_L$ to $2\pi-\theta_L$.
To find the second solution one can make substitution $\bar\theta =
\theta - \pi$ and do similar integration. Result is:
\begin{equation}\label{eq:th_sol}
\vartheta = 2 \tan^{-1}\left[\sqrt{\frac{5}{3}} \tanh \left(\sqrt{\frac{65}{64}}\ \frac{\pm z}{\xi_D}\right)\right],
\qquad
\vartheta = \pi + 2 \tan^{-1}\left[\sqrt{\frac{3}{5}} \tanh \left(\sqrt{\frac{65}{64}}\ \frac{\pm z}{\xi_D}\right)\right].
\end{equation}
The first solution is not topologically protected because it is possible
to move from $-\theta_L$ to $+\theta_L$ by rotating $\bf n$ vector,
always staying in the minimum of spin-orbit interaction. If vector $\bf
n$ is oriented along $\bf\hat z$ for example by magneto-dipolar energy
(not discussed in this text yet) or by spin precession, it will form a
so-called $\bf n$-soliton which has much smaller energy and bigger size.
The second solution is the $\vartheta$-soliton.

$\vartheta$-solitons can exist in HPD state. In this case they have small
core where $\vartheta$ is changing and large tails where $\bf n$ vector
is rotated. This have been discussed in~\cite{1992_mis_hpd_topol} and
numerically calculated in~\cite{2022_zav_theta_hpd}.


\subsection*{Motion of arbitrary soliton (in the rotating frame)} 

Consider a soliton with some distribution $n(z)$ and $\vartheta(z)$ which
are equilibrium solution of Leggett equations.

Consider a soliton which has the same distribution of $n$ and~$\vartheta$ and moves with
velocity~$v$. For the moving soliton we can write~$n(z,dt) = n(z + v\,dt)$ and $ \dot n(z) =
n'\,v$, and same for theta. Now put this into~(\ref{eq:Sr_dndt}) and find
spin $\bf S$ distribution of the moving soliton:
\begin{equation}
\frac{\gamma^2}{\chi_B} {\bf S}(v)
=
\Big[ {\bf n}\ \vartheta' + (1-\ct)\ {\bf n}\times{\bf n'} + \st\ {\bf n'} \Big] v
+ \gamma {\bf H}
+ \omega\ (R{\bf\hat z}-{\bf\hat z})
\end{equation}
Calculate kinetic energy of the soliton using~(\ref{eq:en_m0}), as an
energy difference between moving and static soliton. Note that
expression for energy is same in the rotating frame:
\begin{equation}
E_K = \int_{-\infty}^{\infty} \left[ - ({\bf S}(v) \cdot \gamma {\bf H})
+ \frac{\gamma^2}{2\chi_B}{\bf S}(v)^2
\right] dz
- \int_{-\infty}^{\infty} \left[ - ({\bf S}(0) \cdot \gamma {\bf H})
+ \frac{\gamma^2}{2\chi_B}{\bf S}(0)^2
\right] dz
\end{equation}
$$
= - \frac{\chi_B}{\gamma^2}\ \omega v \int_{-\infty}^{\infty} 2(1-\ct)[{\bf n}\times{\bf n'}]_z\ dz
+ \frac{\chi_B}{\gamma^2}\ \frac{v^2}{2} \int_{-\infty}^{\infty}
\left[
{\bf n}\ \vartheta' + (1-\ct)\ {\bf n}\times{\bf n'} + \st\ {\bf n'}
\right]^2\ dz
$$
The first term is proportional to $\omega\,v$, which is interesting: it
says that if the soliton in the rotating frame has a helical structure
where $\bf n$ rotates around $\bf\hat z$ axis then energy minimum
corresponds to a non-zero velocity. The second term is proportional to
$v^2$, from this term we can get effective mass of the soliton:
\begin{equation}\label{eq:mass}
m = \frac{\chi_B}{\gamma^2}\ \int_z \left[
(\vartheta')^2 + 2(1-\ct)\ ({\bf n'})^2
\right]  dz
\end{equation}
This calculation is valid for small velocities. Arbitrary motion of
$\vartheta$-solitons have been calculated in \cite{1986_rozhkov_solitons}.

\subsection*{Motion of $\vartheta$-solitons} 

Let's look at a 3D case: a soliton perpendicular to $\bf\hat z$ direction
in an experimental cell. It can be considered as a membrane attached to
walls because of some pinning effects. Let's calculate effective mass and
tension of this membrane using profile of the soliton $\vartheta(z)$
from~(\ref{eq:th_sol}).

Tension of the soliton membrane is its total potential energy (spin-orbit + gradient)
per unit area.  If $\bf n$ is uniform the gradient energy~(\ref{eq:en_g0}) is
\begin{equation}
E_{\nabla} = \frac12 \Delta^2 
  \big[(2K_1 + K_2 + K_3) (\nabla\vartheta)^2 - (K_2 + K_3) ({\bf n}\cdot \nabla\vartheta)^2\big].
\end{equation}
For the soliton~(\ref{eq:th_sol0}-\ref{eq:th_sol}) it is
\begin{equation}
F_{\nabla} = \Delta^2 K_1 (\vartheta')^2 = 8 \Delta^2 g_D (\ct + 1/4)^2
\end{equation}
Energy of spin-orbit interaction~(\ref{eq:en_d0}) is also
\begin{equation}
F_{SO} = 8 \Delta^2 g_D (\ct + 1/4)^2
\end{equation}
And tension is integral of total energy density over~$z$:
\begin{equation}
T = 16 \Delta^2 g_D\ \int_{-\infty}^{+\infty}(\ct(z) + 1/4)^2\, dz
\end{equation}
Mass of the soliton per unit area can be found by integrating~(\ref{eq:mass}):
\begin{equation}
M = 8 \frac{\chi_B}{\gamma^2}\frac{g_D}{K_1}
\ \int_{-\infty}^{+\infty} (\ct + 1/4)^2  dz
\end{equation}
Speed of sound is square root of tension divided by mass:
\begin{equation}
C^2 =
2\frac{\Delta^2\gamma^2}{\chi_B} K_1 = 2\cpe^2-\cpa^2
\end{equation}

The first oscillation mode of a circular membrane with radius $R$ is $2.405\,C/R$.

\eject
\printbibliography

\end{document}